\documentclass[aps,prl,twocolumn,amsmath,amssymb,nofootinbib,superscriptaddress,floatfix,reprint,longbibliography]{revtex4-1}

\usepackage{graphicx}% Include figure files
\usepackage{dcolumn}% Align table columns on decimal point
\usepackage{bm}% bold math
\usepackage{tabularx}% Table width
\usepackage{amsmath}% Higher math
\usepackage{textcomp}
\usepackage{upgreek}
\usepackage{color}
\makeatletter

\newcommand{\Rmnum}[1]{\expandafter\@slowromancap\romannumeral #1@}
\makeatother

\usepackage[colorlinks=true, urlcolor=blue, linkcolor=blue, citecolor=blue, pdftex]{hyperref}
\usepackage{ulem} % to strike things out
\usepackage[pdftex]{color} %Titus

\begin{document}

\title{Evolution of Superatomic-Charge-density-wave and Superconductivity under Pressure in AuTe$_2$Se$_{4/3}$}
\author{Xu Chen}
\affiliation{Beijing National Laboratory for Condensed Matter Physics, Institute of Physics, Chinese Academy of Sciences, Beijing 100190, China} \author{Ge Fei}
\affiliation{Laboratory of High Pressure Physics and Material Science (HPPMS), School of Physics and Physical Engineering, Qufu Normal University, Qufu 273100, China}
\author{Yanpeng Song}\affiliation{Beijing National Laboratory for Condensed Matter Physics, Institute of Physics, Chinese Academy of Sciences, Beijing 100190, China}
\author{Tianping Ying}
\email{ying@iphy.ac.cn}
\affiliation{Beijing National Laboratory for Condensed Matter Physics, Institute of Physics, Chinese Academy of Sciences, Beijing 100190, China}
\author{Dajian Huang}\affiliation{Center for High Pressure Science and Technology Advanced Research (HPSTAR), Beijing 100094, China}
\author{Bingying Pan}\affiliation{School of Physics and Optoelectronic Engineering, Ludong University, Yantai 264025, China}
\author{Xiaofan Yang}\affiliation{State Key Laboratory of Surface Physics, Department of Physics, and Laboratory of Advanced Materials, Fudan University, Shanghai 200433, China}
\author{Keyu Chen}\affiliation{Beijing National Laboratory for Condensed Matter Physics, Institute of Physics, Chinese Academy of Sciences, Beijing 100190, China}
\author{Xinhui Zhan}\affiliation{Laboratory of High Pressure Physics and Material Science (HPPMS), School of Physics and Physical Engineering, Qufu Normal University, Qufu 273100, China}
\author{Junjie Wang}
\affiliation{Beijing National Laboratory for Condensed Matter Physics, Institute of Physics, Chinese Academy of Sciences, Beijing 100190, China} \affiliation{School of Physical Sciences, University of Chinese Academy of Sciences, Beijing 100049, China}
\author{Huiyang Gou}
\affiliation{Center for High Pressure Science and Technology Advanced Research (HPSTAR), Beijing 100094, China}
\author{Xin Chen}
\affiliation{Laboratory of High Pressure Physics and Material Science (HPPMS), School of Physics and Physical Engineering, Qufu Normal University, Qufu 273100, China}
\author{Shiyan Li}
\affiliation{State Key Laboratory of Surface Physics, Department of Physics, and Laboratory of Advanced Materials, Fudan University, Shanghai 200433, China}
\author{Jinguang Cheng}
\affiliation{Beijing National Laboratory for Condensed Matter Physics, Institute of Physics, Chinese Academy of Sciences, Beijing 100190, China}
\author{Xiaobing Liu}
\affiliation{Laboratory of High Pressure Physics and Material Science (HPPMS), School of Physics and Physical Engineering, Qufu Normal University, Qufu 273100, China}
\author{Hideo Hosono}
\affiliation{Materials Research Center for Element Strategy, Tokyo Institute of Technology, 4259 Nagatsuta, Midori, Yokohama 226-8503, Japan}
\author{Jian-gang Guo}
\email{jgguo@iphy.ac.cn}
\affiliation{Beijing National Laboratory for Condensed Matter Physics, Institute of Physics, Chinese Academy of Sciences, Beijing 100190, China} \affiliation{Songshan Lake Materials Laboratory, Dongguan, Guangdong 523808, China}
\author{Xiaolong Chen}
\email{chenx29@iphy.ac.cn}
\affiliation{Beijing National Laboratory for Condensed Matter Physics, Institute of Physics, Chinese Academy of Sciences, Beijing 100190, China} \affiliation{School of Physical Sciences, University of Chinese Academy of Sciences, Beijing 100049, China}
\affiliation{Songshan Lake Materials Laboratory, Dongguan, Guangdong 523808, China}

\date{\today}
\begin{abstract}
Superatomic crystal is a class of hierarchical materials composed of atomically precise clusters assembled via van der Waals or covalent-like interactions. AuTe$_2$Se$_{4/3}$, an all-inorganic superatomic superconductor exhibiting superatomic-charge-density-wave (S-CDW), provides a first platform to study the response of their collectively quantum phenomenon to the external pressure in superatomic crystals. We reveal a competition between S-CDW and superconductivity using cutting-edge measurements on thin flakes at low pressures. Prominently, the pressure modulation of S-CDW ordering is 1$\sim$2 order of magnitudes  (0.1 GPa) lower than that of conventional atomic superconductors. As pressure increases to 2.5 GPa, the $T_{\mathrm{CDW}}$ is suppressed and the superconducting transition temperature ($T_{\mathrm{c}}$) is firstly enhanced, and reaches the maximum then quenches with increasing pressure. Above 7.3 GPa, a second superconducting phase emerges, and then a three-fold enhancement in the transition temperature ($T_{\mathrm{c}}$) happens. Analyses of the crystal structure and theoretical calculations suggest a pressure-mediated switch of the conduction channel from the $a$- to the $b$-axis occur, followed by a dimensional crossover of conductivity and the Fermi surface from 2D to 3D.
\end{abstract}
\maketitle

 Interatomic interactions such as van der Waals (vdW) force, ionic, covalent and metallic bonds endow atomic solids versatile but different properties\cite{1,2}. The mesoscopic analogue of atomic solids, coined as ‘superatomic crystal’, are made up of strong-bonded atomic clusters connected by weak interactions or covalent-like quasi-bonding (CLQB) of IIIA-VIA elements\cite{3,4,5}. Known fullerenes\cite{6}, Zintle ions\cite{7}, clusters of aluminum\cite{8}, silicon\cite{9}, boron\cite{10}, and metal chalcogenide molecular clusters\cite{11} are typical examples of superatoms. These superatomic clusters are believed to behave collectively as conventional single atoms with $s$-like, $p$-like, or $d$-like orbitals. Thus, hierarchical designs of the superatomic building blocks and their subsequent stacking sequence provide additional dimension to synthesize novel materials and discover emergent properties\cite{12}.
 
Although the existence of superatoms has long been recognized, the study on superatomic crystals is burgeoning only recently. The electrical conductivity can be greatly altered by varying the mixture of different superatom species\cite{13}. Several interesting properties such as photo-luminescence and electrochemical applications have been discovered\cite{14}. As reported, the majority of superatomic crystals are diamagnetic semiconductors or insulators\cite{11}. Thus far, the mediation of the bonding strength and materials species mainly relies on self-assembling through organic capping ligands\cite{15}. The use of external tuning parameters to control and manipulate the inter-superatomic interactions remains an uncharted territory. The main challenge lies in how to tailor these clusters into crystals and how to probe the impact of inter-cluster-interaction on their properties. 

 \begin{figure*}[tp]
	\includegraphics[clip,width=16.5cm]{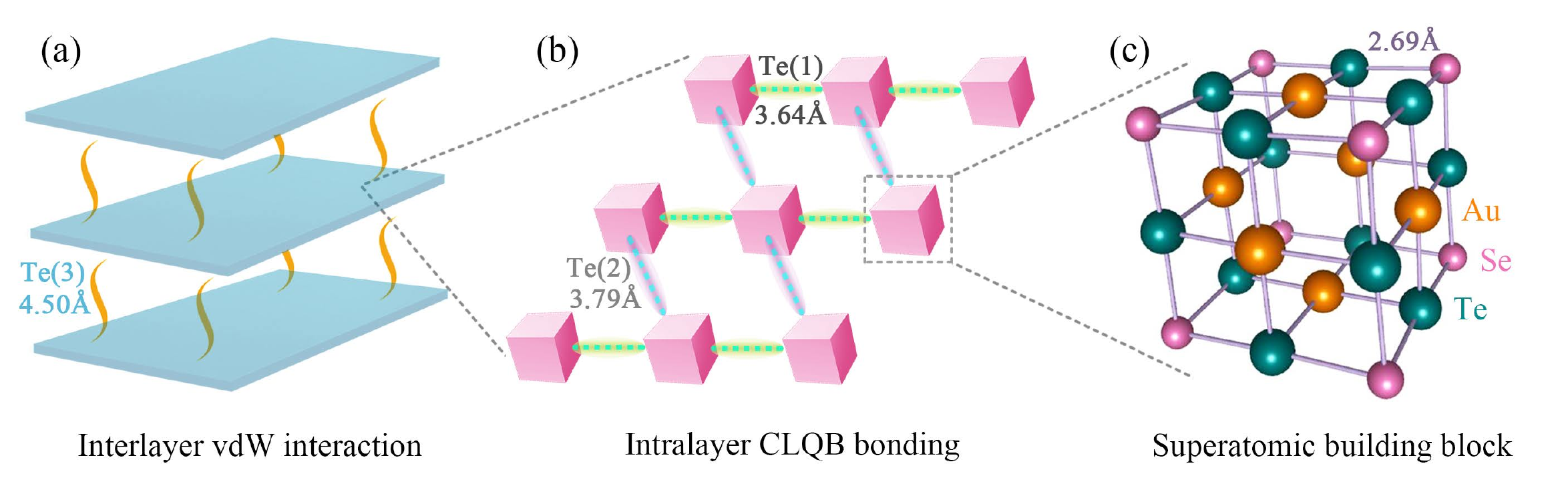}
	\caption{\label{fig1} Structural heterogeneity of superatomic crystal AuTe$_2$Se$_{4/3}$. (a) Each van der Waals (vdW) layer is composed of arrays of woven superatoms. (b) Two covalent-like quasi-bonds of 3.64 $\mathring{A}$ and 3.79 $\mathring{A}$ along $a$-axis and $b$-axis between building blocks. (c) Atomic cube with a composition of Au$_6$Te$_{12}$Se$_8$. Au, Te and Se atoms are represented by golden, cyan and pink balls.}
\end{figure*}
 
Only a handful of all-inorganic superatomic crystals have been reported\cite{16,17}. The newly discovered AuTe$_2$Se$_{4/3}$ superconductor (Fig. \ref{fig1}) is composed of cubic blocks with identical chemical composition of Au$_6$Te$_{12}$Se$_8$\cite{18,19}. Within the cubic, Te atoms are nearest bonded by two Se atoms and four Au atoms to form a rigid cluster, as shown in Fig. \ref{fig1}c. Given a strongly-bonded cluster inside and the weakly CLQB in between clusters, AuTe$_2$Se$_{4/3}$ can be regarded as a superatomic crystal, where the Au$_6$Te$_{12}$Se$_8$ cluster behaves like a superatom. These building blocks are woven together by anisotropic CLQB into 2D sheets with Te-Te1 and Te-Te2 bond length of 3.64 $\mathring{A}$ and 3.79 $\mathring{A}$. The stacking along the $c$-axis through vdW interaction of 4.50 $\mathring{A}$. This material was initially discovered in our lab and later reported to be a natural mineral found in the Koryak Highlands (Russia)\cite{20}. Very recently, scanning tunneling microscope (STM) topographic observations revealed that the electronic states are strikingly different inside and outside the cubes, confirming that the AuTe$_2$Se$_{4/3}$ is a superatomic crystal\cite{21}. Moreover, a superatomic-charge-density-wave (S-CDW) between superatomic blocks along the $b$-axis has been identified because of Te...Te bundle (composed of four Te-Te CLQB)\cite{21}. As we know, CDW and superconductivity, as two kinds of electronic collective states, their relationships are well investigated in many atomic compounds, including IrTe$_2$, ZrTe$_3$, and CsV$_3$Sb$_5$\cite{22,23,24,25,26}. The relationship, however, has remained an uncharted research area in superatomic crystal. 

High-pressure measurement, as a clean and feasible external tuning knob, has been widely used in the realm of atomic solids. In this Letter, we combine the nanofabrication and high-pressure techniques, and for the first time obtain the direct competition between the S-CDW and superconductivity. The S-CDW and superconductivity are sensitive to external pressures and both quenched at a pressure below 2.5 GPa. Superconductivity reenters above 7.3 GPa and $T_{\mathrm{c}}$s are enhanced with increasing applied pressures until 59.7 GPa. The superatomic blocks are found to be intact while experience a crystallographic slip along $b$-axis between 5-10 GPa. Concomitantly, pressure-driven reentrant superconductivity and nonbonding-bonding transition between blocks are observed, which leads to a switch of conducting channel from $a$- to $b$-axis and subsequent 2D to 3D crossover.

\begin{figure*}[tp]
	\includegraphics[clip,width=18cm]{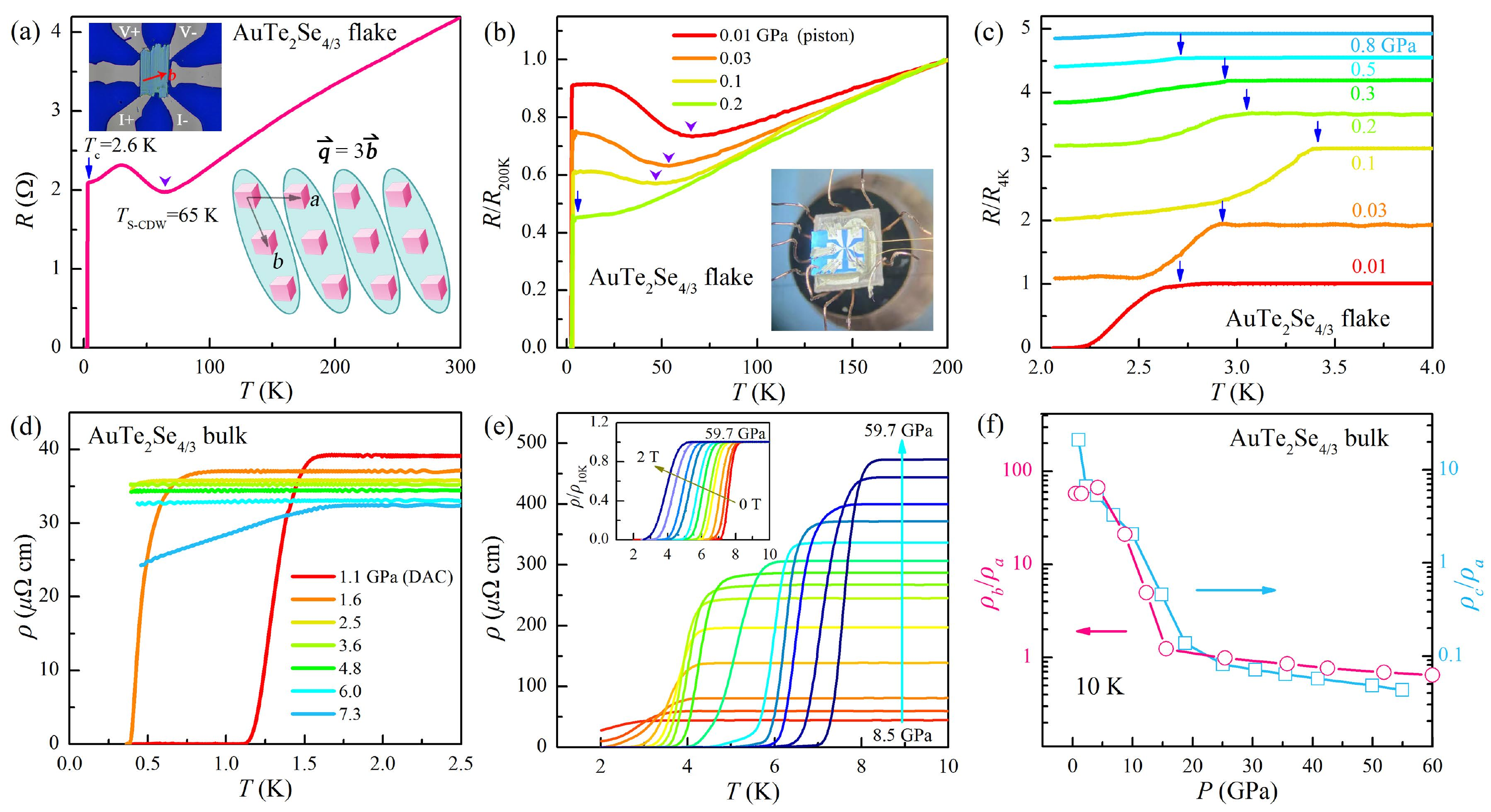}
	\caption{\label{fig2} Evolution of superatomic-charge-density-wave (S-CDW), superconductivity and conducting anisotropic under pressure in AuTe$_2$Se$_{4/3}$. (a) Temperature-dependent resistance ($R$) of AuTe$_2$Se$_{4/3}$ thin flake at ambient pressure. The transition at 65 K (purple arrow) agrees well with the STM data\cite{21}. Blue arrow shows the onset of superconductivity at 2.6 K. Upper inset is the optical image of the fabricated device on flake. Lower inset is the S-CDW form along $b$-axis. (b) and (c) are the temperature-dependent normalized $R$ below 1.0 GPa in different temperature ranges. Inset of (b) is the set-up of device on piston-cylinder apparatus. (d) Temperature-dependent resistivity ($\rho$) of AuTe$_2$Se$_{4/3}$ measured in the range of 1.1 -7.3 GPa using DAC apparatus. (e) $\rho$-$T$ curves at higher pressure up to 59.7 GPa. Reentrance of superconductivity is observed, showing the monotonic increment of normal state $\rho$ (cyan arrow). $\rho$/$\rho_{10K}$ under external magnetic field at 59.7 GPa is shown in the inset. (f) logarithmic plot of resistance ratios ($\rho_{b}$/$\rho_{a}$ and $\rho_{c}$/$\rho_{a}$) as a function of pressure.}
\end{figure*}

 As shown in Fig. \ref{fig2}a, the S-CDW transition is clearly observed at 65 K as AuTe$_2$Se$_{4/3}$ is exfoliated into thin flakes with thickness of $\sim$110 nm, consistent with the STM observation\cite{27,28,29}. We trace the evolution of S-CDW and superconductivity of the thin flake under low pressure by using the Piston-cylinder apparatus, the detailed measurements are shown in Supplementary Materials (SM). It can be seen that the S-CDW is suppressed by external pressure, accompanied by the increase of $T_{\mathrm{c}}$, see Fig. \ref{fig2}b and \ref{fig2}c. As the S-CDW disappears, the $T_{\mathrm{c}}$ reaches its maximum of 3.4 K, then gradually decreases. These behaviors are similar with those in conventional compounds, indicating that superatomic materials can be viewed as a meso-structural version of atomic materials not only from the electronic point of view (shared electrons within the building blocks according to the STM results\cite{21}) but also based on the evolution of electron collective states.

 We further check its transport property under higher-pressures by using diamond anvil cube (DAC) apparatus, see details  in the SM. As shown in Fig. \ref{fig2}d and Fig. S1-S5, the superconducting transition is suppressed above 2.5 GPa, and cannot be observed at 3.6-6.0 GPa down to 0.4 K. Above 7.3 GPa, we observe reemergent superconductivity. Zero resistance is achieved at a pressure of 12.2 GPa (Fig. \ref{fig2}e). The onset $T_{\mathrm{c}}$ increases to 8.5 K at 59.7 GPa, three times higher than the initial $T_{\mathrm{c}}$ of 2.8 K at ambient pressure. Noted that the two superconducting phases are separated by 2.4 GPa. The $\rho$-$T$ curves of the reentrant superconductivity under different magnetic fields are shown in the inset of Fig. \ref{fig2}e, from which we derive its upper critical field $H_{\mathrm{c2}}$ = 3.5 T. Although reentrant superconductivity has been discovered in a variety of compounds, including K$_x$Fe$_{2-y}$Se$_2$ and CsV$_3$Sb$_5$\cite{30,31,32}, this is firstly observed in a superatomic crystal. Figure \ref{fig2}f and Figs. S6-S7 show the pressure-dependent $\rho_{b}$/$\rho_{a}$ and $\rho_{c}$/$\rho_{a}$ at 10 K and other temperatures, respectively. The $\rho_{b}$/$\rho_{a}$ ratios decrease from 65 to $\sim$0.6 when the external pressure is increased from 0.8 to 50 GPa, meanwhile the $\rho_{c}$/$\rho_{a}$ ratios decrease from 11 to $\sim$0.05, suggesting the reentrant  bulk superconductivity is accompanied by conducting channel changes from 2D to 3D. 
 
  \begin{figure*}[tp]
	\includegraphics[clip,width=13cm]{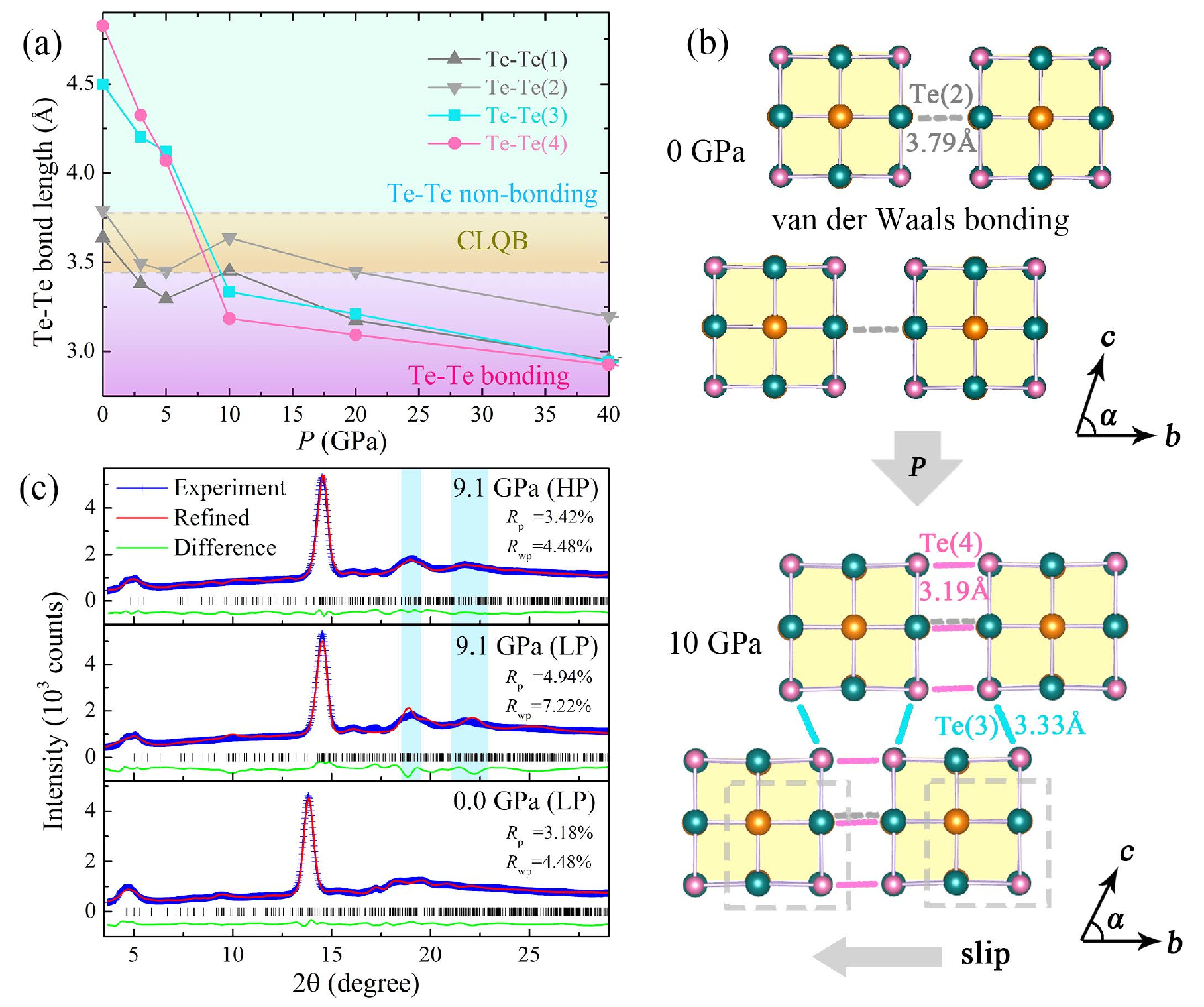}
	\caption{\label{fig3} Structural evolution of superatomic crystal AuTe$_2$Se$_{4/3}$ under external pressure. (a) Pressure-dependence of four Te-Te distances acquired from theoretical calculations.  (b) Illustration of structure transition at ambient pressure and 10 GPa. A slip of the superatomic array along the $b$-axis is schematically drawn. The initial position is depicted in gray dashed squares. Bonds along each crystalline axis is painted in color with reference to (a). We use Te(1) as a shorthand for Te-Te(1) bond and the same goes for other bonds. (c) X-ray diffraction (${\lambda}$ = 0.7107 Å) patterns of AuTe$_2$Se$_{4/3}$ measured at 9.1 GPa and ambient pressure. The green curves are the difference between the observed and refined curves. HP and LP represent high-pressure and low-pressure models that used for the refinements, respectively.}
\end{figure*}
 
Since AuTe$_2$Se$_{4/3}$ can be easily exfoliated into belt-like thin flakes with edges along the $a$ and $b$ directions (Fig. S8), indicating the existence of appreciable in-plane interactions. This type CLQB are characterized by Te...Te bundles along the $a$- and $b$-axes. The Te-Te bondlengths of 3.64 $\mathring{A}$ and 3.79 $\mathring{A}$ lies in this range, while the Te-Te bondlength of 4.5 $\mathring{A}$  beyond it and lies in the range of vdW interaction. We therefore chose Te-Te(2) of 3.8 $\mathring{A}$ as the upper limit of the CLQB interaction. Taking this structure as the initial model, we predicted the structures under high pressures by density functional theory (DFT) calculations. Pressure-dependent inter-cube Te-Te distances, bond angles and lattice constants are plotted in Fig. \ref{fig3}a and Fig. S9. The Te-Te(1) and Te-Te(2) decrease with increasing pressure, and both bondlengths jump a little at 5-10 GPa, but still lie in the CLQB region. Meanwhile, the Te-Te(3) and Te-Te(4) go through a nonbonding-bonding transition and an abrupt lattice collapse. Noted that the pressure of structural transition is qualitatively accordance with the onset pressure at which the second superconducting phase emerges.  
 
 This  nonbonding-bonding transition can be attributed to a slip of the blocks  along the $b$-axis as a response to the applied pressures, as shown in Fig. \ref{fig3}b and Fig. S10. To experimentally confirm these structural changes, we collected 12 diffraction data at pressures from 0 to 24.7 GPa as shown in Fig. S11. We performed the Rietveld refinements by taking the structures calculated by DFT under varying pressures as initial models, and found that the calculated patterns are in good agreement with the experimental ones as shown in Fig. \ref{fig3}c. Here, refinements agreements factor $R_{\mathrm{p}}$ and $R_{\mathrm{wp}}$ converge to below 5\%. The significant changes between the two patterns are the peak shifts, the enhancement of peaks in 2$\theta$ = 18-19 degrees, and the appearance of peaks in 2$\theta$ = 21-23 degrees. In doing so, we were able to extract the lattice parameters (Fig. S12), where the abrupt changes in $a$-, $b$-, and $c$-axes agree with the theoretical predictions in Fig. \ref{fig3}a.

  \begin{figure}[tp]
	\includegraphics[clip,width=8.8cm]{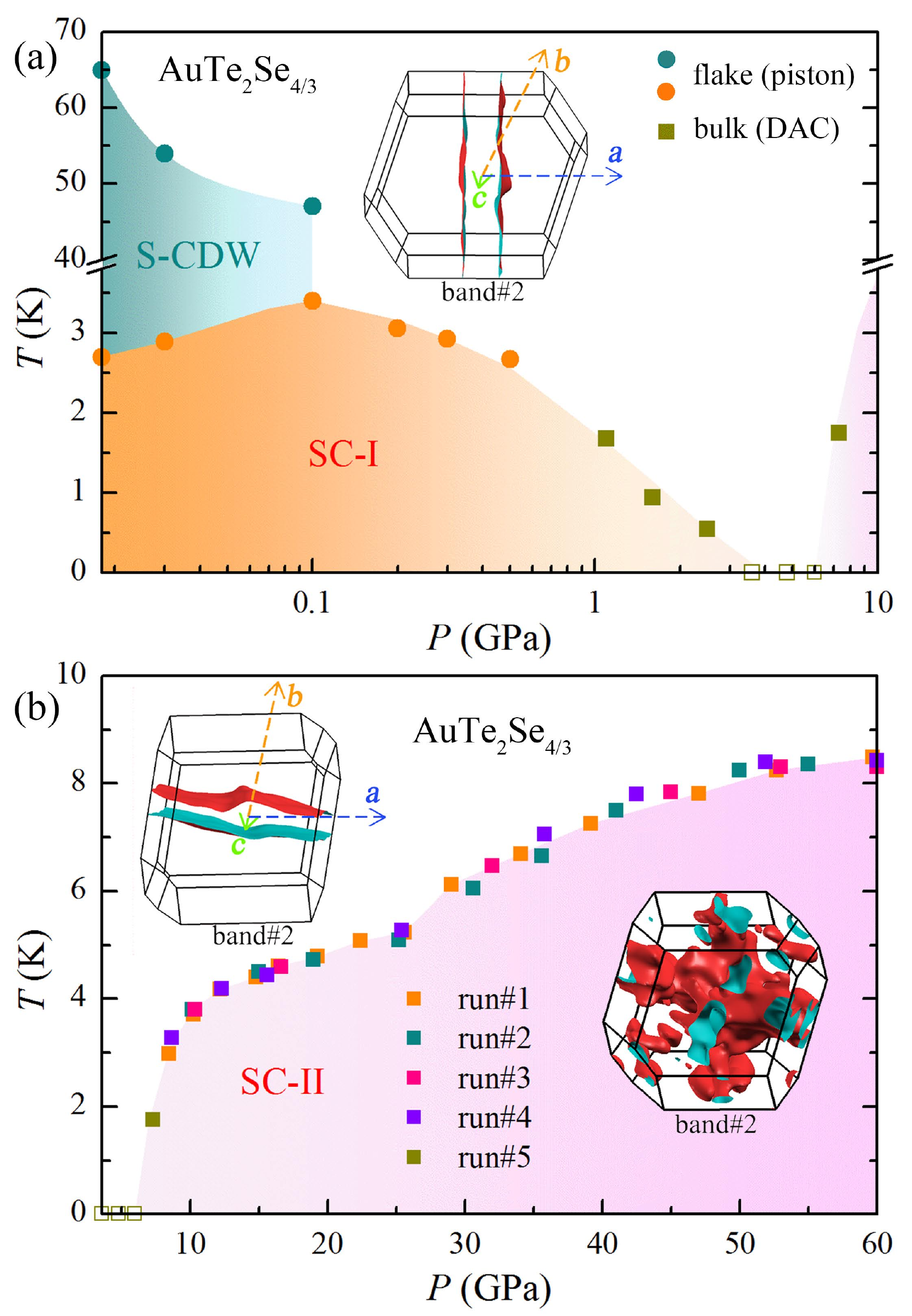}
	\caption{\label{fig4} The S-CDW, superconductivity and Fermi surface topology in pressurized AuTe$_2$Se$_{4/3}$. (a) Phase diagram of AuTe$_2$Se$_{4/3}$ at low pressure. Inset is the dominated band$\#$2 at ambient pressure in first Brillouin zone. (b) Reentrant superconductivity of AuTe$_2$Se$_{4/3}$ at higher pressure region. Insets show the conducting channel along $b$-axis and 3D Fermi surface at 10 GPa and 60 GPa, respectively. Detailed description of the band structure and Fermi surface can be found in SM.}
\end{figure}
 
 To visualize the aforementioned nonbonding-bonding transition, we calculated the electron location function (ELF) at ambient pressure and 10 GPa. 0.5 and 1 are the indicators of covalent and ionic bonds, respectively\cite{33}. A high density of electrons localized between Te and Se atoms within a single Au$_6$Te$_{12}$Se$_8$ cube with an ELF value of 0.6, much higher than  0.11 in between cubes. It means that the superatomic features survive under pressures. At ambient pressure, the inter-cubic Te-Te interactions are mainly along the $a$-axis. At 10 GPa, however, electrons are confined between inter-cubic Te-Te bonds along the $b$-axis, see Fig. S13. Such a charge redistribution along different crystallographic axes is caused by the formation of pressure-induced CLQBs of Te-Te bonding.
 
 The phase diagrams of AuTe$_2$Se$_{4/3}$ over the entire pressure range are plotted in Fig. \ref{fig4}. A competed relationship between S-CDW and superconductivity is clearly mapped out in Fig. \ref{fig4}a. After the first superconductivity is totally suppressed, non-superconducting ground state between 3.6-6 GPa is identified. It is unclear whether there exists a quantum-critical-point between the two superconducting phases. Above 7 GPa, a second superconducting phase shows up, see Fig. \ref{fig4}b. Noted that our $T_{\mathrm{c}}$ in AuTe$_2$Se$_{4/3}$ totally different from the superconducting behavior of elemental Te, see Fig. S14, ruling out the possibility of impurity. Besides, the AuTe$_2$Se$_{4/3}$ does not decompose during measurements, as evidenced by the identical Raman spectra of pristine and pressurized sample, see Fig. S15. For the evolution of Fermi surface, these quasi-2D Fermi surfaces firstly gradually develop in the perpendicular direction with increasing pressures, in line with the superatomic slip of 5-10 GPa. As the pressure further increases, the Fermi surface eventually changes into 3D geometry at 60 GPa \cite{34}. The insets of Fig. \ref{fig4} and Fig. S16-S17 clearly show these variations.

The pressure-dependent structure and transport properties of AuTe$_2$Se$_{4/3}$ exhibit new features distinct from the atomic crystals. The rigid superatomic building blocks can undergo crystallographic slips, resulting in a nonbonding-bonding transition in  inter-cube Te-Te bonds. Upon applying high-enough pressures, a structural crossover from 2D to 3D. In terms of transport property, the anisotropic ratios of $\rho_{c}$/$\rho_{a}$ decrease by three orders of magnitude, confirming the 2D-3D crossover in electronic structure. In addition, the emergence of second superconducting phase and threefold enhancement of $T_{\mathrm{c}}$ should be related to the enhancement of density of states as shown in Fig. S18. Herein, the relationship between structure and transport property is established in superatomic AuTe$_2$Se$_{4/3}$. 
 
  We emphasize that this competition is not a simple resemblance to the atomic CDW ordering. Quantitatively, the superatomic-CDW (S-CDW) is completely suppressed at a pressure of 1$\sim$2 orders of magnitudes (0.1 GPa) lower than that for conventional atomic superconductors (a full list of pressure-quenching CDW thresholds can be found in Table S1). The extremely pressure-sensitive S-CDW suggests it to be a new type of CDW associated with covalent-like quasi-bonding (QLCB) among superatomic clusters. Additionally, the superatomic clusters polarize and anti-polarize periodically in AuTe$_2$Se$_{4/3}$\cite{21}, which leads to the emergence of an anti-polar metallic state that otherwise cannot exist in conventional atomic materials. This means that apart from charge and spin, polarization as another degree of freedom can step in to modulate the properties in a mesoscopic level.
 
 In conclusion, we observe a direct competition between S-CDW and superconductivity at a superatomic level for the first time. The emergence of second superconducting phase is accompanied by relative move of superatomic blocks, rather than changing the geometry and bonding environments of a single block. The whole superatom cluster, therefore, can be treated like a single atom because of the existence of soft CLQB. Remarkably, the discovered S-CDW shows the lowest threshold of quenching the collective ordering than that of CDW in conventional atomic compounds. These results extend our knowledge of the correlation among S-CDW, superconductivity, and structural phase transition in a new class of materials, superatomic compounds. This work significantly modifies physical features in superatomic crystals that have already been identified, as well as will boost the hierarchical design and assembly of novel superatomic superconductors.

~\\
We thank Dr. C.H. Xu from International Centre for Diffraction Data Beijing office for her help on the Rietveld refinement using commercial software Jade Pro.8.6. This work is financially supported by the National Key Research and Development Program of China (No. 2018YFE0202601 and 2017YFA0304700), the National Natural Science Foundation of China (No. 51922105, 11804184, and 11974208), Beijing Natural Science Foundation (Grant No. Z200005), and the Shandong Provincial Natural Science Foundation (ZR2020YQ05, ZR2019MA054, 2019KJJ020). H. H. was supported by a grant from the MEXT Element Strategy Initiative to Form Core Research Center (No.JPMXP0112101001) and JSPS Kakenhi Grants-in-Aid (No.17H06153). \\

\end{document}